\documentclass[10pt,aps, pra, a4paper, notitlepage, twocolumn, amsfonts, amsmath, amssymb,longbibliography]{revtex4-1}
\usepackage[utf8]{inputenc}
\usepackage{graphicx}				
\graphicspath{{Figures/}} 			
\usepackage{epstopdf}
\usepackage{xcolor}
\usepackage{physics}
\usepackage{hyperref}

\newcommand{\ii}{\mathrm{i}}
\newcommand{\ee}{\mathrm{e}}
\newcommand{\intr}[1]{\int\dd{\vb*{r}} \left( #1 \right)}
\newcommand{\rrt}{\left(\vb*{r},t\right)}
\newcommand{\rr}{\left(\vb*{r}\right)}

\begin{document}
\title{
Collective excitations and tunneling dynamics in long bosonic Josephson junctions
}
\author{M. R. Momme}
\author{Y. M. Bidasyuk}
\author{M. Weyrauch}
\affiliation{Physikalisch-Technische Bundesanstalt (PTB), Bundesallee 100, D-38116 Braunschweig, Germany}
\date{\today}

\begin{abstract}
We investigate the low-energy dynamics of two coupled anisotropic Bose-Einstein condensates forming a long Josephson junction. The theoretical study is performed in the framework of the two-dimensional Gross-Pitaevskii equation and the Bogoliubov-de Gennes formalism.
We analyze the excitation spectrum of the coupled Bose condensates and show how low-energy excitations of the condensates lead to multiple-frequency oscillations of the atomic  populations in the two wells. This analysis generalizes the standard bosnic Josephson euqation approach.
We also develop a one-dimensional hydrodynamic model of the coupled condensates, that is capable to reproduce the excitation spectrum and population dynamics of the system.
\end{abstract}

\maketitle

\section{Introduction}

Josephson effects in atomic Bose-Einstein condensates (BECs) receive considerable attention in experimental and theoretical studies as a prominent manifestation of quantum coherence on a macroscopic scale.
These effects have been initially studied for coupled condensates in double-well traps \cite{Zapata1998,Smerzi1997,Raghavan1999,Albiez2005,Levy2007,LeBlanc2011} and coherently coupled spinor condensates \cite{PhysRevA.59.R31,PhysRevLett.105.204101}.
In recent years Josephson effects were also observed and analyzed in more exotic systems such as fermionic superfluids \cite{Valtolina2015}, polariton condensates \cite{abbarchi2013macroscopic}, spin-orbit coupled BECs \cite{Wang2018,PhysRevLett.120.120401} and condensates with attractive interaction \cite{Spagnolli2017}.

A common theoretical description of two coupled BECs is based on the bosonic Josephson equations.
This model has been thoroughly studied theoretically \cite{Smerzi1997,Raghavan1999}.
Various experiments confirm the predictions of this model, which include small-amplitude (plasma) oscillations, large-amplitude anharmonic oscillations and quantum self trapping \cite{Albiez2005,Levy2007,PhysRevLett.105.204101}.
A formal derivation of this model relies on the two-mode description of the system, whereby each of the two condensates retains its density profile and a homogeneous phase.
Such an approximation is valid if the coupling between the two BECs is much weaker than the energy required to create collective excitations inside each condensate.
If this requirement is not fulfilled then internal collective excitations can be generated and  influence the Josephson dynamics of the system.
In order to describe the interference between internal and mutual collective motions of the condensates it is necessary to go beyond the simplified picture of the Josephson equations.
Moreover the structure of collective excitations spectrum and consequently the results of such an interference may be considerably different depending on the geometric properties of the system.
Existing theoretical studies address this question for certain specific types of collective excitations and trap geometries, such as phonon excitations in toroidal condensates \cite{Bidasyuk2016}, and higher modes of the harmonic trap \cite{PhysRevA.89.023614}.
However, various aspects of collective excitations in coupled BECs and their relation to the tunneling transport have yet to be fully explored.

In the present work we analyze the collective excitations and the small-amplitude Josephson oscillations in a system of two highly anisotropic BECs.
Two elongated condensates are placed in parallel to each other forming a so called long Josephson junction (LJJ). Such Josephson junctions were
extensively studied for superconductors \cite{RevModPhys.51.101,PhysRevB.53.6622,PhysRev.164.538} and in atomic BECs mostly in the context of Josephson vortex dynamics \cite{PhysRevA.71.011601,Gil_Granados_2019}.
Our interest in these systems is largely inspired by the experimental results of Ref.~\cite{LeBlanc2011} where the populations of the wells were shown to oscillate at several distinct frequencies, indicating interference with internal collective excitations generated inside each condensate.

In order to describe the Josephson dynamics on the level of collective excitations in the system we need to relate the general physical picture of the Josephson equations with the Bogoliubov-de Gennes formalism. Following the developments in Refs.~\cite{Danshita2005,Paraoanu2001,Burchianti2017} the Josephson plasma oscillations can be related to the lowest dipole mode in the two-well system. Along a similar line of arguments we show here how multiple Bogoliubov modes are involved in the low-energy Josephson dynamics of the anisotropic system.

We model the dynamics of the long bosonic Josephson junction in the mean-field framework of the two-dimensional Gross-Pitaevskii equation. 
The observed dynamical picture is complemented with an analysis of the collective excitation spectrum based on the Bogoliubov-de Gennes approach.
We show how the specific structure of the Bogoliubov modes may lead to various physically relevant effects in the collective dynamics. In addition to enriched Josephson dynamics we identify for a certain range of barrier intensities unexpected quasi-degeneracy of the low-lying Bogoliubov modes with localization of the corresponding quasiparticles at the edges of the junction.
Finally, we develop a simplified one-dimensional hydrodynamic model of the long bosonic Josephson junction, that is able to qualitatively reproduce all important features of the full simulations.

\section{Simulations of long Bose-Josephson junctions}

First we briefly describe simulations of the dynamics of trapped condensed atoms in a bosonic LJJ.
The trap potential  is approximated by an anisotropic harmonic confinement in all spatial dimensions with a Gaussian barrier in $y$ direction, 
\begin{equation}
V(x,y,z) = \frac{m}{2}(\omega_x^2 x^2 + \omega_y^2 y^2 + \omega_z^2 z^2) + V_b \,\ee^{-2 y^2 / \lambda^2}.
\label{eqn:pot}
\end{equation}
The barrier height $V_b$ is a tunable parameter to control the coupling between the two condensate clouds forming in this potential. This setup is patterned similarly to the experimental one described in Ref.~\cite{LeBlanc2011}.
We limit our considerations to values of  $V_b$ larger than the chemical potential $\mu$ of the system such that classical hydrodynamic flow across the barrier is impossible, and the system remains in the tunneling regime.

\begin{figure}[tbp]
	\includegraphics[width=\linewidth]{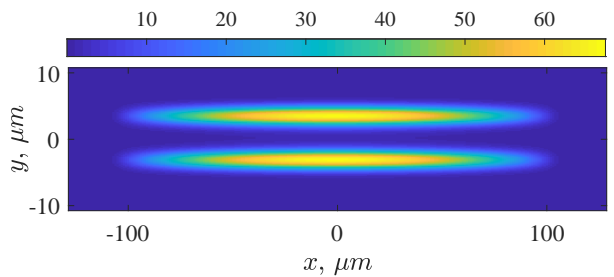}
	\caption{Two-dimensional ground state particle density $\abs{\Psi}^2$ (in arbitrary units) of the long Bosonic Josephson junction. Note the different scales of the $x$ and $y$ axes.}
	\label{fig:config}
\end{figure}

The harmonic part of the potential is assumed to be cigar shaped ($\omega_x \ll \omega_{y,z}$). The barrier cuts the cigar into two parts along the $(x,z)$ plane parallel to the long axis of the trap. In such a configuration the trapping in $z$ direction remains harmonic. Disregarding the dynamics in this direction we may reduce the description of the system to two dimensions using a  Gross-Pitaevskii equation (GPE)
\begin{equation} \label{eqn:GPE}
\ii\hbar\pdv{t}\Psi\rrt = \left(-\frac{\hbar^2 \laplacian}{2m} + V\rr + g \abs{\Psi}^2 \right) \Psi\rrt,
\end{equation}
whith $\vb*{r} = (x,y)$.
The condensate wave function $\Psi\rrt$ is normalized to the total number of atoms in the system $\int \dd{\vb*{r}} \abs{\Psi\rrt}^2 = N$, which is a conserved quantity. The non-linear interaction strength in two dimensions is then given by \cite{Bao2003}
\begin{equation} \label{eqn:interaction_g}
	g = \frac{4\pi\hbar^2 a}{m} \sqrt{\frac{m \omega_z}{2\pi\hbar}},
\end{equation}
where $a$ is the s-wave scattering length and $m$ the mass of a condensate atom.
We here consider  $^{87}\mathrm{Rb}$ atoms with $a = 5.819~\mathrm{nm}$ and $m = 86.91~\mathrm{u}$ and
choose  $\omega_x = 2\pi\cdot 1~\mathrm{Hz}$, $\omega_y = \omega_z = 2\pi\cdot 50~\mathrm{Hz}$, a barrier $1/e^2$ half-width $\lambda = 3.5~\mu\mathrm{m}$, and a total number of atoms  $N = 2.0 \cdot 10^4$. However, most of our results  should be qualitatively similar for any bosonic LJJ in the tunneling regime.
For the calculations presented in this section  we use a barrier amplitude $V_b/h = 375~\mathrm{Hz}$.
The ground state stationary solution of the GPE (\ref{eqn:GPE}) shows the shape of two nearly one-dimensional parallel  condensates (see Fig.~\ref{fig:config}). We find a  chemical potential $\mu/h= 273~\mathrm{Hz}$ and consequently $V_b / \mu = 1.37$.
Note that the chemical potential is still considerably larger than the trapping frequencies. Therefore, the mean-field treatment based on the GPE (\ref{eqn:GPE}) is justified.

Josephson effects in bosonic systems lead to particle transfer between the wells. The relative population imbalance $Z(t)$ is the dynamical quantity which we use to quantify these effects
\begin{equation}\label{eqn:Zdef}
Z(t) = \frac{N_1(t) - N_2(t)}{N},
\end{equation}
where $N_1(t)$ and $N_2(t)$ denote the number of atoms in each well ($N_1(t) + N_2(t) = N$). These populations are related to the solutions $\Psi\rrt$ of the GPE. They are defined as integrals of the particle density over each well
\begin{equation}\label{eqn:N12def}
N_j(t) = \int_{A_j}\dd{\vb*{r}} \abs{\Psi\rrt}^2,
\end{equation}
where $j=1,2$ and $A_j$ is the area subtended
 by well $j$.

The second dynamical quantity of interest is the phase difference between the condensates
\begin{equation}
	\varphi(t) = \theta_2(t) - \theta_1(t),
\end{equation}
where each phase is defined by averaging over corresponding well
\begin{equation}
	\theta_j(t) = \arg{\int_{A_j}\dd{\vb*{r}}\Psi\rrt}.
\end{equation}
One may show that this quantity is canonically conjugate to the population imbalance.

The time evolution of the population imbalance and relative phase of two BECs is often accurately described by the Josephson equations \cite{Smerzi1997,Raghavan1999}
\begin{align}
&\dot{Z} = - J\sqrt{1-Z^2} \sin(\varphi), \label{eqn:TMM_sym_Zdot}\\
&\dot{\varphi} = \Lambda Z + J \frac{Z}{\sqrt{1-Z^2}} \cos(\varphi). \label{eqn:TMM_sym_phasedot}
\end{align}
Here, $\Lambda$ describes the intra-well interactions and $J$ characterizes the coupling between the two condensates. These quantities may be expressed in terms of the solutions of the GPE. A derivation of the Josephson equations with explicit definitions for $\Lambda$ and $J$ based on the two-mode model may be found in the \hyperref[app:TMM]{Appendix}. The ratio between $\Lambda$ and $J$ defines the physical region of validity of this model. It is expected to be valid in the region $1 \ll \Lambda/J \ll N^2$ which is also called  `Josephson regime' \cite{PitaevskiiStringari2016,Leggett2001}.
The value of this ratio is strongly dependent on the barrier height. For the range of configurations considered in the present study this ratio ranges from $\Lambda/J \approx 7$ for ($V_b/\mu=1$) to $\Lambda/J \approx 9800$ for ($V_b/\mu=2$) which is well inside the Josephson regime.
The Josephson equations (\ref{eqn:TMM_sym_Zdot},\ref{eqn:TMM_sym_phasedot}) have been successfully applied to describe various double-well experiments \cite{Albiez2005,Levy2007,PhysRevLett.105.204101}. Here we will use these equations as a basic reference and analyze their limitations when applied to anisotropic condensates.

In the case of a small initial population imbalance $\abs{Z_0}\ll 1$ or initial phase difference $\varphi_0\ll\pi/2$, the equations (\ref{eqn:TMM_sym_Zdot},\ref{eqn:TMM_sym_phasedot}) reduce to a harmonic oscillator equation with the frequency
\begin{equation}\label{eqn:plasma_frequency}
\omega_p = \sqrt{J(\Lambda + J)},
\end{equation}
known as plasma frequency.
Thus the model predicts harmonic oscillations for small initial imbalances or phase differences. The value of the plasma frequency marks another important limitation of the model based on the system (\ref{eqn:TMM_sym_Zdot},\ref{eqn:TMM_sym_phasedot}). In order for the Josephson oscillations to be decoupled from the internal oscillations inside each condensate it is necessary that $\omega_p \ll \omega_{x,y,z}$ \cite{PitaevskiiStringari2016}.
This condition is not fulfilled for our system and therefore we can expect the effects of intra-well collective excitations to influence the Josephson dynamics.

For larger initial population imbalance or phase difference the Josephson equations predict the development of anharmonic large amplitude oscillations and the transition to the self-trapped regime at the critical population imbalance $Z_{cr}=2\sqrt{J/\Lambda(1-J/\Lambda)}$ \cite{Raghavan1999}.

\begin{figure}[tbp]
	\begin{center}
		\includegraphics[width=\linewidth]{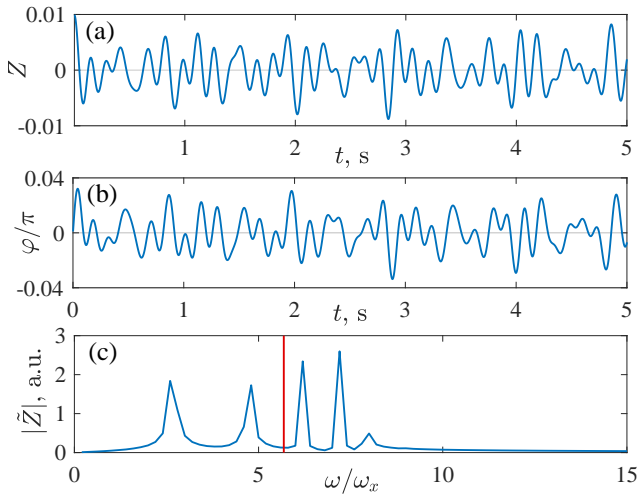}
		\caption{(a) Population imbalance $Z(t)$ and (b) relative phase $\varphi(t)$ as a function of time extracted from a GPE simulation. (c) The Fourier spectrum $|\tilde{Z}(\omega)|$ of the population imbalance (in arbitrary units). The vertical red line indicates the plasma frequency $\omega_p$ predicted by the two-mode model.}
		\label{fig:Zsequences}
	\end{center}
\end{figure}

In order to study the near-equilibrium dynamics of the junction we perform numerical simulations using the GPE (\ref{eqn:GPE}). The initial state is prepared close to the ground state with an initial imbalance $Z_0 = 0.01$ inside of the plasma oscillation regime.
In a real experiment such an initialization can be realized with an additional linear offset potential that is switched off at $t=0$.
In our simulations the system evolves freely for $5~\mathrm{s}$ and the population imbalance $Z(t)$ is determined at equally spaced times $t_m$.
The observed time series  $Z(t)$ is shown in the upper panel of Fig. \ref{fig:Zsequences} revealing clearly population oscillations not determined by a single frequency.

\begin{figure}[tbp]
	\begin{center}
		\includegraphics[width=\linewidth]{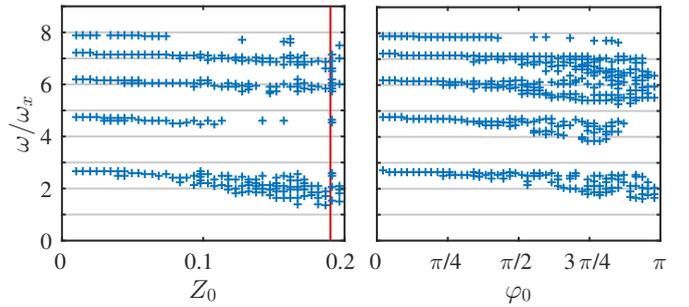}
		\caption{Frequencies corresponding to peaks in the Fourier spectrum $\tilde{Z}$ of the population imbalance as a function of the initial imbalance $Z_0$ (left) and the initial phase difference $\varphi_0$ (right). The vertical red line on the left panel indicates the critical population imbalance $Z_{cr} = 0.19$ for the onset of the self-trapping regime predicted by the two-mode model.}
		\label{fig:InitialZ}
	\end{center}
\end{figure}

We analyze the spectral properties of the oscillations by taking the discrete Fourier transform (DFT) of the time series $Z(t_m)$
\begin{equation}
	\tilde Z(\omega_n) = \sum_{m=0}^{M} Z(t_m) \ee^{-\ii \omega_n t_m}, \quad \omega_n=2\pi n/t_M,
\end{equation}
where $M$ is the total number of time steps in the simulation.
The Fourier-spectrum $\tilde Z(\omega)$ shows several well separated peaks (see Fig.~\ref{fig:Zsequences}, lower panel). We consider then $Z(t)$ to be a superposition of several harmonic oscillations. We extract the frequencies of these distinct oscillations simply by locating the maxima in the spectrum. It is worth noticing, that none of the detected frequencies can be related to the plasma frequency predicted by the two-mode model.
In order to investigate this finding further, we vary the initial population imbalance and initial phase difference. The results plotted in Fig.~\ref{fig:InitialZ} show that multiple peaks in the spectrum can be observed at arbitrarily small initial imbalance or phase difference. Their positions remain stable in the regions $Z_0\ll Z_{cr}$ and $\varphi_0\ll\pi/2$. Furthermore, the frequencies get smaller with increasing $Z_0$ or $\varphi_0$ as is similarly the case for the plasma frequency \cite{Raghavan1999,Martinez2018}. Finally, in the regime of large-amplitude oscillations distinct frequencies can hardly be distinguished and the DFT spectrum does not provide useful information. In the rest of the paper we will only consider the regime of small oscillations with $Z_0 = 0.01$ and $\varphi_0 = 0$, where the frequencies are well defined and stable.

\section{Bogoliubov-de-Gennes formalism}

\begin{figure*}[tbp]
	\includegraphics[width=\linewidth]{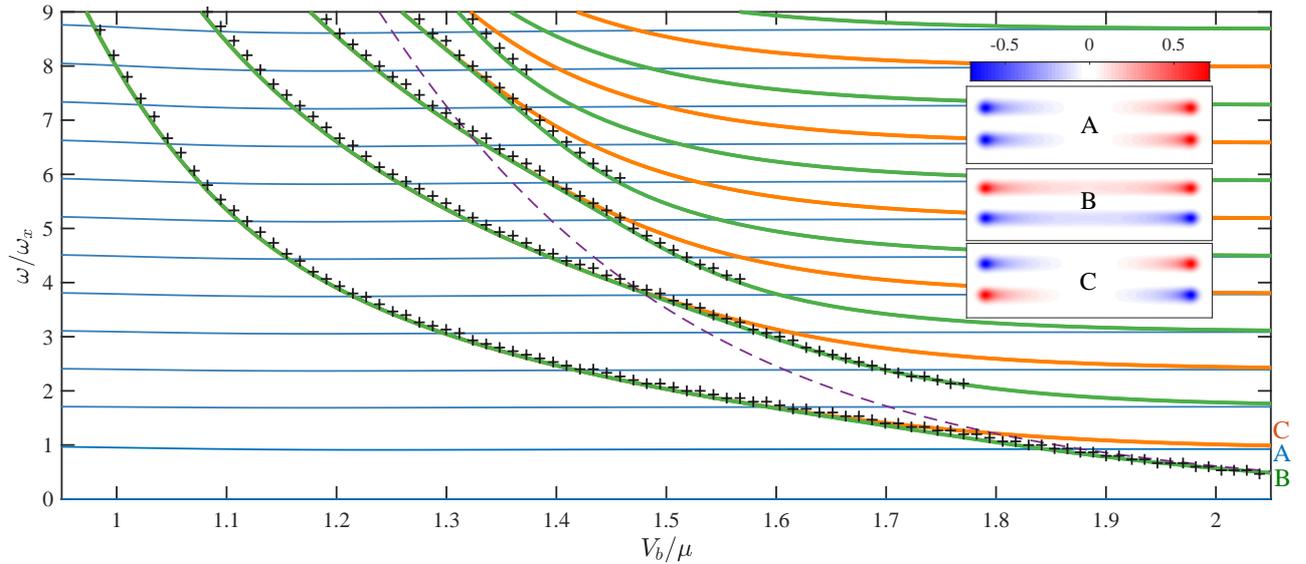}
	\caption{Spectrum of elementary excitation of the long Josephson junction. Solid lines of different colors correspond to the Bogoliubov modes of different symmetry: A --- transversely symmetric (blue lines), B --- transversely antisymmetric but longitudinally symmetric (green lines) and C --- antisymmetric in both directions (red lines). The insets show the spatial distribution of the lowest excitations of each of the three types (scale of the color bar is in arbitrary units). The corresponding modes are marked on the main plot with the letter on the right. Black crosses correspond to the peak positions in the Fourier spectrum $\tilde Z(\omega)$ obtained from the full GPE simulations. The purple dashed line is the estimate of the Josephson plasma frequency $\omega_p$ based on the two-mode model.
	}
	\label{fig:BarrierScan}
\end{figure*}

The results of the previous section show that several modes with different frequencies contribute to the population oscillations in the system. Furthermore, the number of these modes and their frequency is almost independent of the oscillation amplitude. This indicates that non-linear mode mixing effects are negligible in the considered regime. Nevertheless, such a multi-mode population oscillation spectrum is a clear indication that intra-well collective excitations influence the tunneling dynamics.
In order to further analyze how different collective excitations influence the Josephson dynamics we employ the Bogoliubov-de Gennes formalism \cite{PitaevskiiStringari2016,Danshita2005}. We first write the condensate wave function in the form
\begin{equation}\label{eqn:perturbed_gs}
\Psi\rrt = \ee^{-\frac{\ii\mu t}{\hbar}} \left[\psi_0\rr + \delta\psi\rrt \right].
\end{equation}
where $\psi_0$ is the stationary ground state, $\mu$ is the corresponding chemical potential and $\delta\psi\rrt$ is a perturbation of the form
\begin{equation} \label{eqn:Bogoliubov_ansatz}
\delta\psi\rrt = \sum_k c_k \left[ \ee^{-\ii\omega_k t} u_k\rr + \ee^{\ii\omega_k t} v_k^{\ast}\rr \right],
\end{equation}
with the additional condition that the perturbation is small ($c_k \ll \sqrt{N}$).
After inserting this ansatz into the GPE (\ref{eqn:GPE}) and considering only the terms linear in $c_k$ we obtain the familiar system of Bogoliubov-de Gennes equations
\begin{equation}\label{eqn:BdG}
\begin{aligned}
	\hbar\omega_k u_k = (\hat{H}_0 + 2g \abs{\psi_0}^2 -\mu) u_k +  g \psi_0^2 v_k, \\	
	-\hbar\omega_k v_k = (\hat{H}_0 + 2g \abs{\psi_0}^2 -\mu) v_k +  g \psi_0^2 u_k,
\end{aligned}
\end{equation}
where
\begin{equation}
\hat{H}_0 = -\frac{\hbar^2\laplacian}{2m} + V\rr.
\end{equation}
The usual normalization condition for the Bogoliubov modes reads
\[
\int d\mathbf{r} \left( |u_k|^2 - |v_k|^2 \right) = 1.
\]
This linear system of equations can be solved numerically in order to obtain eigenmodes and their corresponding frequencies $\omega_k$. We solve the system (\ref{eqn:BdG}) for different values of the barrier height $V_b$. The calculated spectrum as a function of $V_b$ is presented in Fig.~\ref{fig:BarrierScan}. We also perform full GPE simulations with barrier heights in the same range to compare the frequencies extracted from the time series $Z(t)$ with the Bogolibov spectrum.

As one can see in Fig.~\ref{fig:BarrierScan} only some of the Bogoliubov modes correlate with the spectrum of population oscillations.
In order to understand this fact let us see how the linear excitations of the form (\ref{eqn:Bogoliubov_ansatz}) enter into the dynamics of the population imbalance of the two-well system.
To this end we introduce the following operator representation of the population imbalance
\[
Z = \frac{1}{N} \langle \Psi | \hat Z | \Psi \rangle, \qquad
\hat Z\rr = \left\{\begin{aligned} 1,\quad & \mathbf{r}\in A_1 (y<0),\\-1,\quad & \mathbf{r}\in A_2 (y>0).\end{aligned}\right.
\]
which is valid for any wave function $\Psi$. We can now insert here the wave function of the form (\ref{eqn:perturbed_gs},\ref{eqn:Bogoliubov_ansatz}).
We keep only terms up to first order in $c_k$ and assume (without loss of generality) that $\psi_0\rr$, $u_k\rr$ and $v_k\rr$ are purely real functions. Taking also into account that $\langle \psi_0 |\hat Z| \psi_0 \rangle=0$ the final expression yields
\begin{equation}\label{eq:zt_bdg}
	Z(t) = \frac{2}{N} \sum_k c_k D_k \cos(\omega_k t).
\end{equation}
Here
\begin{equation}
\label{eq:d2_bdg}
D_k = \langle \psi_0 | \hat Z | u_k + v_k \rangle
\end{equation}
is the excitation amplitude of the $k$'th Bogoliubov mode by an operator $\hat Z$. In other words the set of coefficients $D_k$ essentially characterizes the condensate response to the small perturbation proportional to $\hat Z$ \cite{PitaevskiiStringari2016,PhysRevA.56.587}.

The equation (\ref{eq:zt_bdg}) shows that a populated Bogoliubov mode with index $k$ results in a transfer of population with the frequency $\omega_k$, but only if the coefficient $D_k$ is non-zero.
To identify which Bogoliubov modes may correspond to non-zero values of $D_k$ we should consider symmetry properties of these modes. Both the potential (\ref{eqn:pot}) and the ground state are symmetric with respect to axes reflections in longitudinal ($x$) and transverse ($y$) directions. Therefore the solutions of (\ref{eqn:BdG}) are expected to possess certain reflection symmetries in these directions as well. In Fig.~\ref{fig:BarrierScan} we distinguish three subsets of levels (plotted with different colors) based on their symmetry properties. These states are (A) transversely symmetric, (B) transversely antisymmetric but longitudinally symmetric and (C) antisymmetric in both directions. In the following we will refer to these states as A, B and C for shorter notation.

From the structure of Eq.~(\ref{eq:d2_bdg}) one can predict that $D_k$ may be non-zero only if $u_k+v_k$ is symmetric in $x$ and antisymmetric in $y$. Consequently the states of type B are the only subset of modes that can possibly contribute to the Josephson dynamics identified by $Z(t)$.
This argument is confirmed by the results in Fig.~\ref{fig:BarrierScan} showing a nearly perfect match of the spectra extracted from GPE dynamics with the Bogoliubov modes of type B.

\begin{figure}[tbp]
	\includegraphics[width=\linewidth]{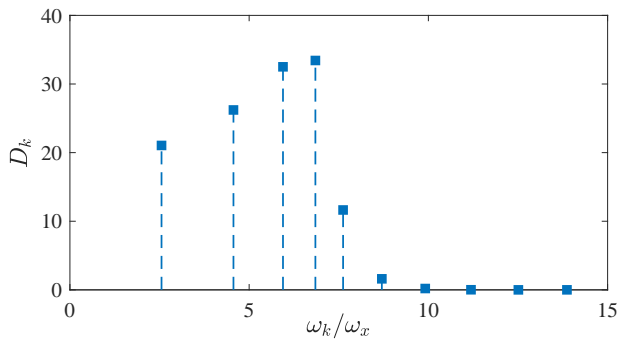}
	\caption{Excitation amplitudes $D_k$ for the 10 lowest collective excitations of the type B calculated with the barrier height $V_b/h = 375\,\mathrm{Hz}$.}
	\label{fig:dkcoefs}
\end{figure}

The above symmetry arguments provide necessary conditions for any of the collective excitations to be traceable in the population imbalance $Z(t)$. However, these conditions are by no means sufficient. The judgement on the number of non-zero amplitudes $D_k$ can be done only by calculating them numerically or measuring them. Such calculations show that there is always only a finite number of essentially non-zero values in this set. In Fig.~\ref{fig:dkcoefs} we show the values of $D_k$ and $\omega_k$ corresponding to the lowest type B excitations for the same barrier heights used in the previous section. Apparently only the five lowest excitations show amplitudes $D_k$ significantly larger than zero. One may also notice a similarity of Fig.~\ref{fig:dkcoefs} and the spectrum $\tilde Z(\omega)$ presented in the lower panel of Fig.~\ref{fig:Zsequences}. This shows that the spectrum of population oscillations can in general be predicted by analyzing the Bogoliubov modes of the system.
This also means that the number of observable modes is a property of the system, and does not depend on the initial perturbation imposed in the numerical simulation. 

\section{General properties of the Bogoliubov spectrum}

We now investigate further the complex structure of the calculated excitation spectrum.
The first obvious observation is that the barrier affects differently the excitations of different symmetries. In particular, the excitations of type A which are symmetric in the transverse direction are practically unaffected by the changes of the barrier height. These solutions are node-less in $y$ and represent purely longitudinal excitations. They originate from the collective excitations of a single one-dimensional BEC and are well approximated by a simple analytical formula based on hydrodynamic approximation \cite{Menotti2002} (see Fig.~\ref{fig:CompBdG} for direct comparison)
\begin{equation}\label{eq:1d-sym-spectrum}
\omega = \omega_x \sqrt{\frac{k(k+1)}{2}},
\end{equation}

\begin{figure}[tbp]
	\includegraphics[width=\linewidth]{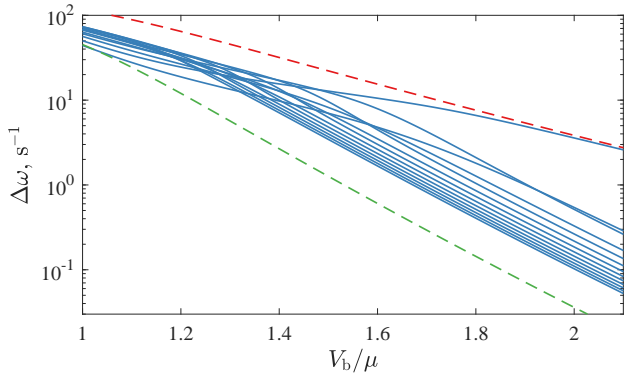}
	\caption{Level spacings $\Delta\omega = \omega_k^{(B,C)} - \omega_k^{(A)}$ between the levels of type A and corresponding levels of types B and C. The plasma frequency $\omega_p$ (red dashed line) and coupling strength $J$ (green dashed line) are shown for comparison. Note the logarithmic scale of the vertical axis, which means that a straight line on the figure corresponds to an exponential decay of the curve.}
	\label{fig:levelsplit1}
\end{figure}

The modes of types B and C are strongly affected by the barrier height but asymptotically they converge to the symmetric counterparts and are expected to match them exactly in the limit of two completely uncoupled condensates. In Fig.~\ref{fig:levelsplit1} we show the level spacings $\Delta\omega = \omega_k^{(B,C)} - \omega_k^{(A)}$ between the levels of type A and corresponding levels of types B and C.
We see that in the high barrier limit all level spacings decay exponentially with growing $V_b/\mu$. Interestingly, all the decay rates are related to parameters defined in the two-mode model. The lowest pair of levels converge at the rate of plasma frequency $\omega_p$ but all the other pairs converge at the same rate as the coupling parameter $J$.
In the limit of a very high barrier $V_b/\mu > 1.9$ where $\omega_p < \omega_x$, higher modes disappear from $Z(t)$ dependence and the lowest oscillation frequency converges to the estimated value of the plasma frequency $\omega_p$ (see Fig.~\ref{fig:BarrierScan}). This indicates that validity of the Josephson equations (\ref{eqn:TMM_sym_Zdot},\ref{eqn:TMM_sym_phasedot}) is restored in this region.

\begin{figure}[tbp]
	\includegraphics[width=0.5\linewidth]{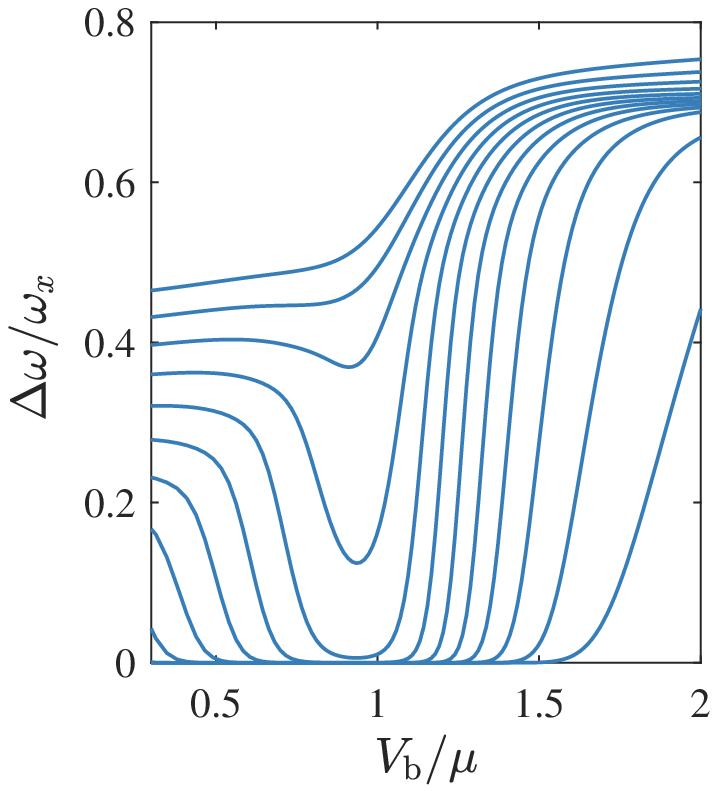}\includegraphics[width=0.5\linewidth]{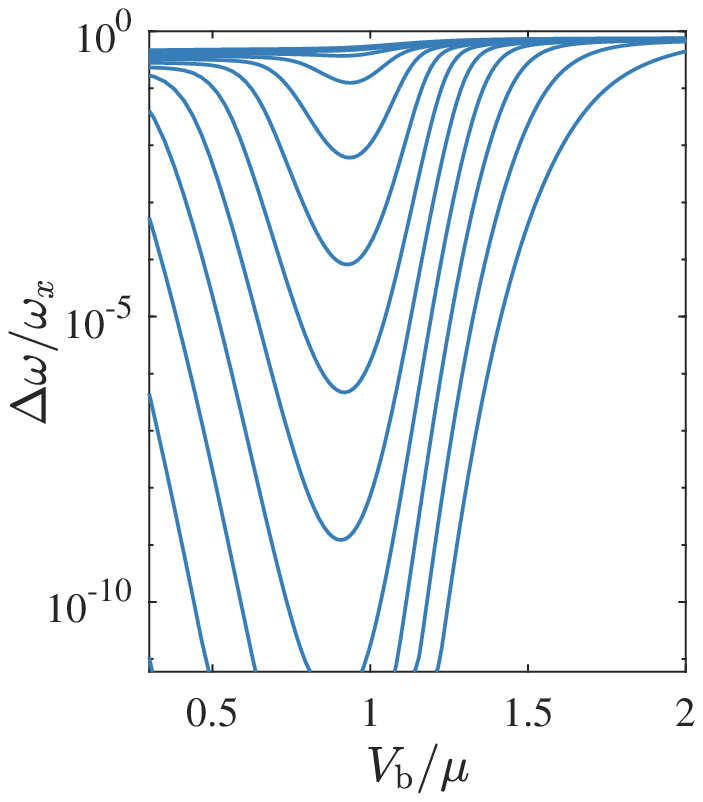}
	\caption{Level spacings $\Delta\omega = \omega_k^{(C)} - \omega_k^{(B)}$ shown in linear (left panel) and logarithmic (right panel) scale.}
	\label{fig:levelsplit2}
\end{figure}

\begin{figure}[tbp]
	\includegraphics[width=\linewidth]{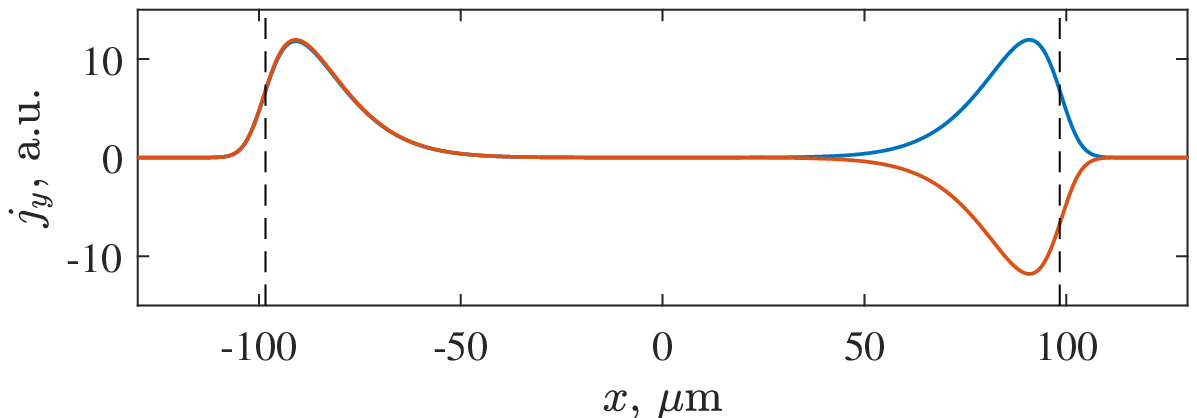}
	\includegraphics[width=\linewidth]{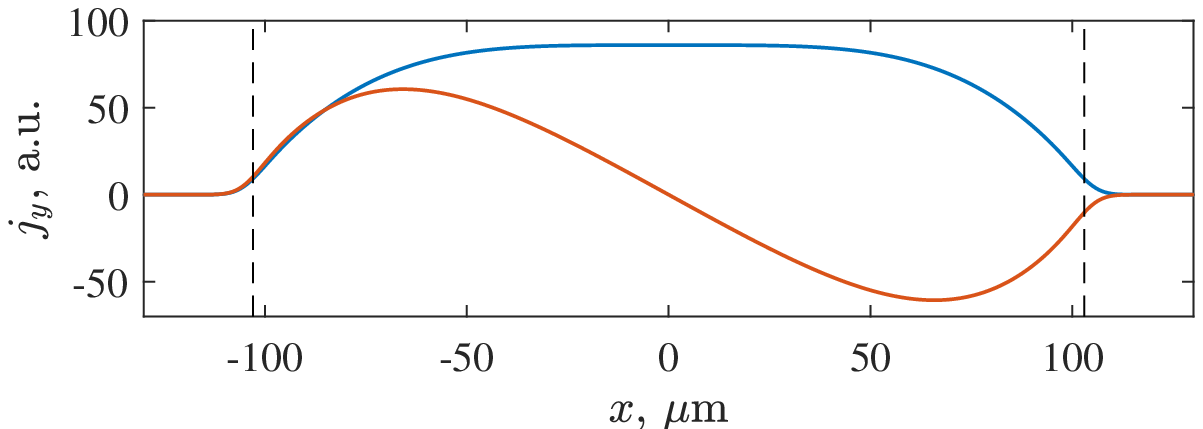}
	\caption{Tunneling current density distribution corresponding to the lowest Bogoliubov excitations of type B (blue lines) and C (red lines). Top panel corresponds to barrier height $V_b/h = 325 \mathrm{Hz}$ ($V_b/\mu \approx 1.26$), bottom panel --- $V_b/h = 600 \mathrm{Hz}$ ($V_b/\mu \approx 1.93$). Vertical dashed lines on both panels show the Thomas-Fermi boundary of the system.}
	\label{fig:current}
\end{figure}

Levels of types B and C also show an unexpected degeneracy in the region of intermediate barrier heights. The spacings between these levels are shown in Fig.~\ref{fig:levelsplit2}. We show them in both linear and logarithmic scales to highlight that for all of the affected levels the minimal level spacing is observed at the same barrier height around $V_b/\mu \approx 1$. This quasi-degeneracy arises due to coupling between two condensates.
In order to understand the origin of this degeneracy we look into the distribution of the tunneling current initiated by these excitations. The transverse tunneling current density can be written in the following form

\[
J_y(x,t) = \frac{\hbar}{2 m \ii} \int dy \left( \Psi^* \partial_y \Psi - \Psi \partial_y \Psi^* \right).
\]
Inserting here the wave function (\ref{eqn:perturbed_gs},\ref{eqn:Bogoliubov_ansatz}) we get
\[
J_y(x,t) = \sum_k c_k j_y^{(k)}(x) \sin(\omega_k t),
\]
where
\[
j_y^{(k)}(x) = \frac{\hbar}{m} \int dy \left[ \psi_0 \partial_y (u_k - v_k) - (u_k - v_k) \partial_y \psi_0 \right]
\]
is the characteristic current density associated with $k$'th Bogoliubov mode. These current densities for the lowest Bogoliubov modes of types B and C are shown in Fig.~\ref{fig:current}. We see that in the region of quasi-degeneracy the tunneling current in these modes is localized at the edges of the junction and two edge-localized excitations basically decouple from each other.
A similar effect was also described for the superconducting LJJs \cite{PhysRev.164.538} and attributed to the Meissner effect in the junction.
One may also see from Fig.~\ref{fig:BarrierScan}, that these edge-localized states are always involved in the dynamics of population imbalance $Z(t)$ (the crosses cover all regions of quasi-degeneracy).

\section{Effective one-dimensional model of elementary excitations}
So far all of our descriptions for the bosonic LJJ have been in two spatial dimensions. However, the geometry of the system suggests that we can consider it as two coupled one-dimensional systems.
For the symmetric modes the one-dimensional treatment of the system is quite simple and leads to the analytical expression (\ref{eq:1d-sym-spectrum}). However, for the spectrum of antisymmetric modes to the best of our knowledge no analytical expressions exist. The model developed in Ref.~\cite{Abad2013} describes dispersion relations of coupled infinite homogeneous 1D condensates, so while it captures some essential properties of the spectra, it can not reproduce the observed level degeneracy and edge-localized states.
In the following we develop a simplified one dimensional hydrodynamic model that is still able to reproduce qualitatively the behavior of antisymmetric Bogoliubov modes.
Our aim is to reduce the system to two coupled one-dimensional condensates. The reduction scheme that we propose is similar in spirit to the usual two-mode model.
It relies on the additional assumption that the dimensions can be separated. So the long axis of the trap $x$ does not couple to the short axis $y$.
In each well the wave functions can be written as a product state and the total wave function reads
\begin{align}
\Psi(x,y,t) = \Psi_1(x,t) \chi_1(y) + \Psi_2(x,t) \chi_2(y).
\end{align}
We assume the solutions in $y$ dimension $\chi_{1,2}(y)$ to be time independent. Additionally we define $\chi_1$ and $\chi_2$ such that they are real, orthogonal and normalized to unity. Our trapping potential is obviously separable $V(x,y)/\hbar = \tilde{V}_x(x) + \tilde{V}_y(y)$.
Inserting this ansatz into the GPE (\ref{eqn:GPE}) leads to a system of two coupled one-dimensional equations
\begin{align}\label{eqn:1DC-GPE}
&\ii \pdv{t} \Psi_1(x,t) = \left(-\frac{\hbar}{2m} \partial_x^2 + \tilde{V}_x + g_{1D} \abs{\Psi_1}^2 \right) \Psi_1 \notag\\
&- \left(K + F(\abs{\Psi_1}^2 + \abs{\Psi_2}^2)\right) \Psi_2 - F(\Psi_1^\ast \Psi_2 + \Psi_2^\ast \Psi_1) \Psi_1 \notag\\
&+ M \left(\abs{\Psi_2}^2 \Psi_1 + (\Psi_1^\ast \Psi_2 + \Psi_2^\ast \Psi_1) \Psi_2\right),
\end{align}
where for the second component the indices are interchanged. The coefficients that appear in these equations are defined as follows
\begin{align}
g_{1D} &= \frac{g}{\hbar} \int \dd{y} \chi_1^4, \\
K &= - \int\dd{y} \left(-\frac{\hbar}{2m} \chi_1 \partial_y^2 \chi_2 + \chi_1 \tilde{V}_y \chi_2 \right), \\
F &= - \frac{g}{\hbar} \int\dd{y} \chi_1^3 \chi_2, \\
M &= \frac{g}{\hbar} \int\dd{y} \chi_1^2 \chi_2^2.
\end{align}
In the Josephson tunneling regime the overlap between functions $\chi_1$ and $\chi_2$ is small and we can safely assume $M \ll F \ll g_{1D}$. Then the equations simplify to
\begin{align}\label{eqn:1DC-GPE-red}
&\ii \pdv{t} \Psi_1(x,t) = \left(-\frac{\hbar}{2m} \partial_x^2 + \tilde{V}_x + g_{1D} \abs{\Psi_1}^2 \right) \Psi_1 \notag\\
&- \left(K + F(\abs{\Psi_1}^2 + \abs{\Psi_2}^2)\right) \Psi_2, \notag\\
&\ii \pdv{t} \Psi_2(x,t) = \left(-\frac{\hbar}{2m} \partial_x^2 + \tilde{V}_x + g_{1D} \abs{\Psi_2}^2 \right) \Psi_2 \notag\\
&\shoveright{- \left(K + F(\abs{\Psi_1}^2 + \abs{\Psi_2}^2)\right) \Psi_1.}
\end{align}

These equations permit a one-dimensional description of the double-well system. They are similar to the commonly used model for two-component condensates with coherent coupling. However they contain additional terms proportional to $F$ which in general cannot be neglected. A more detailed study of the effects of these additional terms would require a separate study. We only mention here that $K$ is in general not sign-definite and neglecting the nonlinear terms proportional to $F$ may lead to unphysical results. This problem is usually rectified by taking $K$ as a phenomenological positive definite parameter or imposing additional approximations (see e.g. Ref.~\cite{PhysRevA.81.025602} for a discussion and references).

In order to further simplify our model, we recall that for the parameters considered in the present work the 1D system is deeply in the Thomas-Fermi regime. This suggests that a hydrodynamic approximation may be valid to describe the system. We rewrite the equations (\ref{eqn:1DC-GPE}) with the usual Madelung decomposition $\Psi_1(x,t) = \sqrt{\rho_1} \ee^{\ii\theta_1}$, where $\rho_1(x,t)$ is the density and $\theta_1(x,t)$ is the phase, which gives
\begin{align}\label{eqn:Hydro}
\dot{\rho}_1 = \;& -\frac{\hbar}{m} \partial_x\left(\rho_1 \partial_x \theta_1 \right) \notag\\
& - 2 \left(K + F(\rho_1 + \rho_2)\right) \sqrt{\rho_1 \rho_2} \sin(\theta_2 - \theta_1), \notag\\
\dot{\theta}_1 = \;& - \frac{\hbar}{2m} \left(\partial_x \theta_1\right)^2 - \tilde{V}_x - g_{1D}\rho_1 \notag\\
& +\left(K + F(\rho_1 + \rho_2)\right)\sqrt{\frac{\rho_2}{\rho_1}} \cos(\theta_2-\theta_1),
\end{align}
where we neglected the quantum pressure terms of the form $\partial_x(\sqrt{\rho})/\sqrt{\rho}$.
The equations of the other component have interchanged indices.

To study the antisymmetric low energy excitations we consider a stationary state for each well and we add a perturbation with an opposite sign in each well:
\begin{align}\label{eq:hydro-ansatz}
&\rho_1(x,t) = \rho_0(x) + \delta\rho(x,t), && \theta_1(x,t) = -\tilde{\mu} t - \delta\theta(x,t), \notag\\
&\rho_2(x,t) = \rho_0(x) - \delta\rho(x,t), && \theta_2(x,t) = -\tilde{\mu} t + \delta\theta(x,t),
\end{align}
where $\tilde{\mu} = \mu/\hbar$. The Thomas-Fermi ground state density is
\begin{equation}\label{eq:TF-GS}
\rho_0(x) = \frac{\tilde{\mu}-\tilde{V}_x(x)+K}{g_{1D}}
\end{equation}
in the region where $\tilde{\mu}+K>V_x$ and zero otherwise.
With this ansatz we linearize the coupled equations,
\begin{align}
\pdv{t} \delta\rho = &\; \frac{\hbar}{m} \partial_x \left(\rho_0 \partial_x \delta\theta \right) - 4 \left(K + 2F \rho_0\right) \rho_0 \delta\theta,\\
\pdv{t} \delta\theta = &\; \left(g_{1D}  + \frac{K}{\rho_0}\right) \delta\rho.
\end{align}
Combining the two equations we get one second-order equation for $\delta\rho$
\begin{align}
\pdv[2]{t}\delta\rho = \;& \frac{\hbar}{m} \partial_x \left((g_{1D}\rho_0 + K) \partial_x \delta\rho - K \frac{\delta\rho\, \partial_x \rho_0}{\rho_0} \right) \notag\\
& - 4\left(K + 2F\rho_0\right) \left(g_{1D}\rho_0 + K\right) \delta\rho.
\end{align}
The term proportional to $(\partial_x\rho_0)/\rho_0$ can be safely neglected in the Thomas-Fermi regime. We finally assume the solutions to be periodic in time $\delta\rho(x,t) = \cos(\omega t) \delta\rho(x)$, which leads to the following equation
\begin{align} \label{eqn:Hydro1Dfinal}
-\omega^2 \delta\rho = \;& \frac{\hbar}{m} \partial_x \left((g_{1D}\rho_0 + K) \partial_x \delta\rho  \right) \notag\\
& - 4\left(K + 2F\rho_0\right) \left(g_{1D}\rho_0 + K\right) \delta\rho.
\end{align}

This equation provides a hydrodynamic approximation for the transversely antisymmetric collective excitations of the system. 
Using the explicit expression for the Thomas-Fermi ground state density (\ref{eq:TF-GS}) this equation can be easily recast in terms of the chemical potential and the trap potential, which makes it numerically tractable. In some limiting cases the equation (\ref{eqn:Hydro1Dfinal}) can be solved analytically. In particular, in the limit of uncoupled condensates ($K \rightarrow 0, F \rightarrow 0$) it is straightforward to show that the spectrum (\ref{eq:1d-sym-spectrum}) is reproduced as is expected in this limit. Also if one considers the excitations of the same shape as the ground state ($\delta\rho(x,t) = Z(t) \rho_0$) then the result of the two-mode model is recovered and after integrating out the spatial dimension we get a single frequency $\omega = \omega_p$ identical to Eq.~(\ref{eqn:plasma_frequency}).

\begin{figure}[tbp]
	\includegraphics[width=\linewidth]{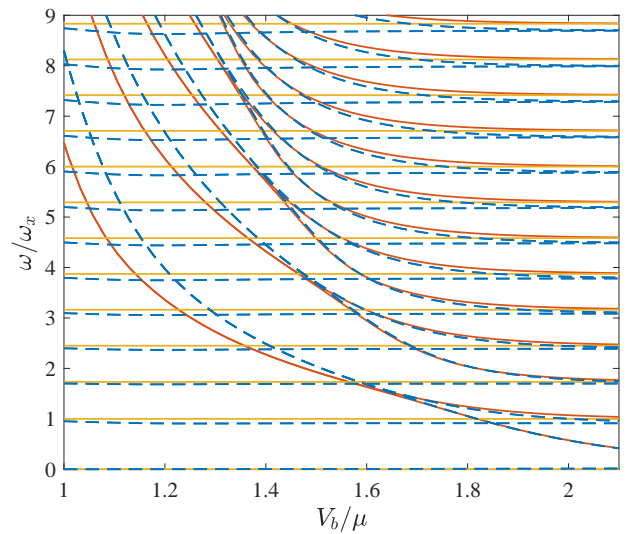}
	\caption{Spectrum of collective excitations in the hydrodynamic approximation based on the equations (\ref{eq:1d-sym-spectrum}) (solid yellow lines) and (\ref{eqn:Hydro1Dfinal}) (solid red lines) compared to the full solution of the Bogoliubov-de Gennes equations (\ref{eqn:BdG}) (dashed blue lines).
	}
	\label{fig:CompBdG}
\end{figure}

In figure \ref{fig:CompBdG} we compare numerical solutions of equation (\ref{eqn:Hydro1Dfinal}) with the Bogoliubov spectrum of the two-dimensional system. The parameters $g_{1D}$, $K$ and $F$ were estimated based on the ground state GPE solutions. The two approaches show reasonable agreement, which is better in the high barrier region. Specific features of the spectrum, such as formation of quasi-degenerate pairs of levels is also reproduced in the developed hydrodynamic approximation. This result justifies the general validity of this approach and applicability of the derived equation (\ref{eqn:Hydro1Dfinal}).

\begin{figure}[tbp]
	\includegraphics[width=\linewidth]{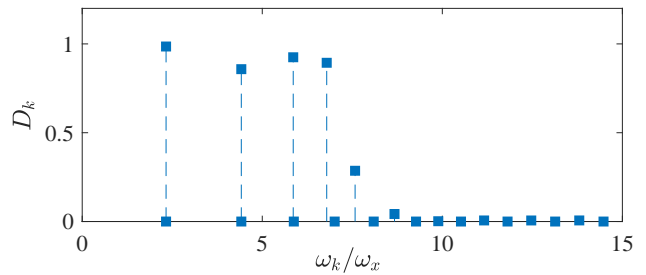}
	\caption{Excitation amplitudes $D_k$ in the hydrodynamic approximation (in arbitrary units). The barrier height is $V_b/h = 375\,\mathrm{Hz}$.}
	\label{fig:dkcoefs_hydro}
\end{figure}

Using the ansatz (\ref{eq:hydro-ansatz}) we can also write the population imbalance in the form analogous to (\ref{eq:zt_bdg})
\[
Z = \frac{2}{N} \sum_k D_k \cos(\omega_k t),
\]
where the coefficients $D_k$  have a similar meaning to the excitation amplitudes discussed in the previous section and are defined by the solutions of Eq.~(\ref{eqn:Hydro1Dfinal}) as follows
\[
D_k = \int dx \delta\rho_k.
\]
The values of these coefficients are shown in Fig.~\ref{fig:dkcoefs_hydro}. We see that they also qualitatively reproduce the results in Figs.~\ref{fig:dkcoefs} and~\ref{fig:Zsequences}. A number of additional zero-valued coefficients on the figure correspond to excitations of type C, which are also included in the spectrum of Eq.~(\ref{eqn:Hydro1Dfinal}).

\section{Conclusions}

In the present work we  investigated the low-energy dynamics of long bosonic Josephson junctions.
In the simulations based on the time-dependent Gross-Pitaevskii equation we observe
oscillations of the population imbalance consisting of multiple well-defined frequencies, that persist even for arbitrarily small initial perturbation.

In order to understand this behavior we analyze the spectrum of elementary excitations obtained from the Bogoliubov-de Gennes equations.
By analyzing the condensate response to the imposed population imbalance we find the Bogoliubov modes that can contribute to the population transfer. This allows us to explain and predict the multi-mode spectrum of population oscillations observed in the dynamical simulations.

From the general structure of the excitations spectrum in the region of intermediate barrier heights we discover the development of quasi-degeneracy of low-lying levels. We connect this phenomenon with localization of corresponding modes at the edges of the junction. The geometry of LJJ allows such edge-localized excitations to decouple and form degenerate pairs. Quite interestingly, these edge-localized oscillations are also always observed in the dynamics of the population imbalance.

We also developed an effective one-dimensional model of two coupled condensates, that combines a hydrodynamic approximation in the longitudinal direction and quantum tunneling in the transverse direction of the barrier. This simplified model is shown to qualitatively reproduce all essential features of the excitation spectrum obtained using the Bogolubov-de-Gennes analyis:
The excitation frequencies are in good agreement in a wide range of barrier heights. The model also correctly predicts multiple frequencies for the population dynamics of the condensate.

\appendix*

\section{The two-mode model}\label{app:TMM}
The two-mode model is a common approach to describe Josephson effects in a bosonic double-well system. We review it here to introduce the concepts and establish a notation.
The central idea of the two-mode model is that the total wave function can be described with only two modes. Each one is (roughly) localized in one well and has one coherent phase
\begin{equation} \label{eqn:two_mode_ansatz}
\Psi\rrt = \sqrt{N_1(t)}\: \ee^{\ii\theta_1(t)} \phi_{1}\rr + \sqrt{N_2(t)}\: \ee^{\ii\theta_2(t)} \phi_{2}\rr.
\end{equation}
$N_{1,2}$ is the occupation, $\theta_{1,2}$ is the phase and $\phi_{1,2}$ are the time-independent wave functions of each mode, which have unit norm and are orthogonal. One can obtain them with the standard ansatz of combining the ground state solution $\phi_g\rr$ with the first excited state that has a node in the barrier region $\phi_{ex}\rr$:
\begin{equation} \label{eqn:def_of_phione_phitwo}
\phi_{1}\rr = \frac{\phi_{g}\rr - \phi_{ex}\rr}{\sqrt{2}}, \quad \phi_{2}\rr = \frac{\phi_{g}\rr + \phi_{ex}\rr}{\sqrt{2}}.
\end{equation}
The time-dependent variables of the two-mode model are the population imbalance $Z(t)$ and the relative phase $\varphi(t)$,
\begin{equation}
Z(t) = \frac{N_1(t) - N_2(t)}{N}, \qquad \varphi(t) = \theta_2(t) - \theta_1(t),
\end{equation}
where $N$ is the total number of atoms with $N_1+N_2 = N$. Putting the ansatz (eqn. \ref{eqn:two_mode_ansatz}) into the GPE leads to the well-known two-mode model equations
\begin{align}
&\dot{Z} = - J\sqrt{1-Z^2} \sin(\varphi) + I (1-Z^2) \sin(2\varphi), \label{eqn:TMM1}\\
&\dot{\varphi} = \Lambda Z + J \frac{Z}{\sqrt{1-Z^2}} \cos(\varphi)  - I Z \left(2+\cos(2\varphi)\right),\label{eqn:TMM2}
\end{align}
where the parameters are defined as
\begin{align}
\Lambda = &\; \frac{g N}{\hbar} \intr{\phi_1^4},\label{eqn:LambdaDef}\\
J = &\; -\frac{2}{\hbar} \intr{\frac{\hbar^2}{2m} \grad{\phi}_1 \grad{\phi}_2 + \phi_1 \phi_2 V + gN\phi_1^3 \phi_2},\\
I = &\; \frac{g N}{\hbar} \intr{\phi_1^2 \phi_2^2}.
\end{align}
Here $\Lambda$ is the on-site interaction, $J$ describes the coupling and $I$ is called interaction tunneling (see \cite{Ananikian2006}).
In case of small oscillations $\abs{Z} \ll  1$ and $\varphi \ll \pi/2$ the equations reduce to a harmonic oscillator equation with the frequency
\begin{equation}
\omega_{p} = \sqrt{\Lambda J - 2\Lambda I + J^{2} - 5JI + 6I^{2}}.
\end{equation}
This is called plasma frequency and it characterizes the population transfer between the wells. In most configurations $\Lambda \gg J \gg I$, so $I$ can be neglected and one obtains 
$\omega_p = \sqrt{J(\Lambda + J)}$ as quoted in Eq.~\ref{eqn:plasma_frequency}.

Results of experiments show that the two-mode model provides better predictions if the parameters are modified. Especially $\Lambda$ is usually too large with the definition \ref{eqn:LambdaDef}. This is due to the approximation that the wave functions $\phi_1\rr$ and $\phi_2\rr$ are time-independent. Relaxing that condition by allowing the wave functions to depend on the population of the respective wells like this $\phi_j\rr \rightarrow \phi_j(\vb*{r},N_j(t))$ for $j=1,2$, provides a more accurate representation of the physics. This ansatz was implicitly introduced in Refs. \cite{Jezek2013, Nigro2017} and it was termed effective two-mode model. We provide a slightly different (and to our opinion more transparent) derivation.
The ansatz for the total wave function is
\begin{equation}
	\Psi\rrt = \sqrt{N_1} \phi_1(\vb*{r},N_1) \ee^{\ii\theta_1} + \sqrt{N_2} \phi_2(\vb*{r},N_2) \ee^{\ii\theta_2}.
\end{equation}
Here $N_1,N_2,\theta_1$ and $\theta_2$ are time-dependent.
This can be seen as an adiabatic approximation because the wave function for one well is assumed to be in the ground state at the corresponding population.
If we put this ansatz into the GPE (\ref{eqn:GPE}) we obtain the equations
\begin{align}
\dot{Z} = &\; - \left(J + \tilde{J} Z \right) \sqrt{1-Z^2} \sin(\varphi) + I (1-Z^2) \sin(2\varphi), \label{eqn:ETMM1}\\
\dot{\varphi} = &\; \Delta E + \Lambda Z + J \frac{Z}{\sqrt{1-Z^2}} \cos(\varphi) \notag\\
& + \tilde{J} \frac{2Z^2 -1}{\sqrt{1-Z^2}} \cos(\varphi) - I Z \left(2+\cos(2\varphi)\right), \label{eqn:ETMM2}
\end{align}
where $J,\tilde{J},I,\Delta E$ and $\Lambda$ are functions of $Z$
\begin{align}
\Delta E(Z) = &\; \frac{1}{\hbar}\intr{\frac{\hbar^{2}}{2m} \abs{\grad{\phi_1}}^2 + \phi_1^2 V + \frac{g N}{2} \phi_1^4}\notag\\
&\; - \frac{1}{\hbar}\intr{\frac{\hbar^{2}}{2m} \abs{\grad{\phi_2}}^2 + \phi_2^2 V + \frac{g N}{2} \phi_2^4}, \\
\Lambda(Z) = &\; \frac{g N}{2\hbar} \intr{\phi_1^4 + \phi_2^4},\\
J(Z) = &\; -\frac{2}{\hbar} \int\dd{\vb*{r}} \bigg( \frac{\hbar^2}{2m} \grad{\phi}_1 \grad{\phi}_2 + \phi_1 \phi_2 V \notag\\
&\; + \frac{g N}{2} \left(\phi_1^3 \phi_2 + \phi_1 \phi_2^3 \right) \bigg) ,\\
\tilde{J}(Z) = &\; \frac{g N}{\hbar} \intr{\phi_1 \phi_2^3 - \phi_1^3 \phi_2},\\
I(Z) = &\; \frac{g N}{\hbar} \intr{\phi_1^2 \phi_2^2},
\end{align}
because $\phi_1(\vb*{r},Z)$ and $\phi_2(\vb*{r},Z)$ depend on it. For $Z=0$ the two-mode model parameters are recovered.
The relation $\Lambda \gg J \gg I$ still holds, so $I$ can be neglected safely. Also the changes in $J$ and $\tilde{J}$ are expected to be small so $J(Z) \approx J(0) = J$ and $\tilde{J}(Z) \approx \tilde{J}(0) = 0$. With that the equations of motion are
\begin{align}
\dot{Z} = &\; - J \sqrt{1-Z^2} \sin(\varphi),\\
\dot{\varphi} = &\; \Delta E(Z) + \Lambda(Z)\cdot Z + J \frac{Z}{\sqrt{1-Z^2}} \cos(\varphi).
\end{align}
The key of the effective model is to approximate the functions $\Delta E(Z)$ and $\Lambda(Z)$. However at this point it is not clear if and how they are related. For that let us introduce the local chemical potential of each well
\begin{equation}
	\mu_j = \intr{\frac{\hbar^2}{2m} \abs{\grad{\phi_j}}^2 + V\rr \phi_j^2 + g N_j \phi_j^4}.
\end{equation}
The difference between the local chemical potentials is indeed
\begin{equation}
	 \mu_1 - \mu_2 = \Delta E + \Lambda Z.
\end{equation}
This relation clearly shows that $\Delta E$ and $\Lambda$ cannot be treated separately.
A Taylor approximation at $Z=0$ gives
\begin{equation}
	\mu_1 - \mu_2 = \eval{(\mu_1 - \mu_2)}_{Z=0} + \eval{\left(\dv{\mu_1}{Z} - \dv{\mu_2}{Z}\right)}_{Z=0} Z + \order{Z^2}.
\end{equation}
The first term has to vanish due to the symmetry of the double well. The linear term can be changed to contain the appropriate populations $N_1$ and $N_2$ for each well
\begin{equation}
	\eval{\left(\dv{\mu_1}{Z} - \dv{\mu_2}{Z}\right)}_{Z=0} = \frac{N}{2} \left( \eval{\dv{\mu_1}{N_1}}_{N_1=\frac{N}{2}} + \eval{\dv{\mu_2}{N_2}}_{N_2=\frac{N}{2}} \right)
\end{equation}
Again the symmetry of the double-well causes the derivatives of $\mu_1 $ and $\mu_2$ to be the same, so we finally get 
\begin{align}
	\dot{Z} = &\; - J \sqrt{1-Z^2} \sin(\varphi),\\
	\dot{\varphi} = &\; \Lambda_{\mathrm{eff}}\cdot Z + J \frac{Z}{\sqrt{1-Z^2}} \cos(\varphi),
\end{align}
where
\begin{equation}
	\Lambda_{\mathrm{eff}} = N \eval{\dv{\mu_1}{N_1}}_{N_1=N/2}.
\end{equation}
This derivative can be obtained by calculating the ground state $\phi_1$ for several different populations $N_1$ and putting it in the functional $\mu_1$. We use this definition of the parameter $\Lambda$ for all two-mode model estimates presented in the present work. 

\bibliographystyle{unsrt}
\bibliography{Library2}

\end{document}